\documentclass[11pt,letterpaper,notoc]{JHEP3}
\usepackage{amsmath,cite,feynmf}
\input epsf.tex

\newcommand{\beq}{\begin{eqnarray}}
\newcommand{\eeq}{\end{eqnarray}}
\newcommand{\nn}{\nonumber}
\def\ltap{\ \raise.3ex\hbox{$<$\kern-.75em\lower1ex\hbox{$\sim$}}\ }
\def\gtap{\ \raise.3ex\hbox{$>$\kern-.75em\lower1ex\hbox{$\sim$}}\ }

\def\mpl{M_{\rm Pl}}

\newcommand\openone{\leavevmode\hbox{\small1\normalsize\kern-.33em1}}

\newcommand{\lsim}{\,\raise.3ex\hbox{$<$\kern-.75em\lower1ex\hbox{$\sim$}}\,}
\newcommand{\gsim}{\,\raise.3ex\hbox{$>$\kern-.75em\lower1ex\hbox{$\sim$}}\,}
\newcommand{\be}{\begin{eqnarray}}
\newcommand{\ee}{\end{eqnarray}}

\newcommand{\bg}{ {\mathbf g}}
\newcommand{\ba}{ {\mathbf a}}
\newcommand{\GC}{GC}
\renewcommand{\zeta}{\varepsilon}

\title{Effective Field Theory for Massive
Gravitons and Gravity in Theory Space}

\author{Nima Arkani-Hamed, Howard Georgi and Matthew D. Schwartz\\
Jefferson Laboratory of Physics, Harvard
  University, Cambridge, MA 02138 \\ email:
  \email{arkani@carnot.harvard.edu, georgi@physics.harvard.edu,matthew@feynman.princeton.edu}}

\preprint{HUTP-02/A051\\}

\abstract{
We introduce a technique for restoring general coordinate invariance
into theories where it is explicitly broken. This is the analog for
gravity of the Callan-Coleman-Wess-Zumino formalism for gauge theories.
We use this to elucidate the properties of interacting 
massless and massive gravitons. For a single graviton with a
Planck scale $\mpl$ and a mass
$m_g$, we find that there is a sensible 
effective field theory 
which is valid up to a 
high-energy cutoff $\Lambda$ parametrically above $m_g$. 
Our methods allow for a transparent understanding of the 
many peculiarities associated with massive gravitons, among them 
the need for the Fierz-Pauli form of the Lagrangian, the presence or 
absence of the 
van Dam-Veltman-Zakharov discontinuity in general backgrounds,
and the onset of non-linear effects and the breakdown of the effective
theory at large distances from heavy sources. The natural sizes 
of all non-linear
corrections beyond the Fierz-Pauli term are easily determined.  
The cutoff scales as 
$\Lambda \sim (m_g^4 \mpl)^{1/5}$ 
for the Fierz-Pauli theory, but can be raised to 
$\Lambda \sim (m_g^2 \mpl)^{1/3}$ in certain non-linear extensions. 
Having established that these models make sense as effective theories, there
are a number of new avenues for exploration, 
including model building with gravity in theory space
and constructing gravitational dimensions.}

\begin{document}
\begin{fmffile}{grav12pics}

\section{Introduction}
It has recently been realized that non-gravitational extra dimensions can
be generated dynamically from fundamentally four-dimensional gauge theories
\cite{Arkani-Hamed:2001ca, Hill:2000mu, Cheng:2001vd}.
These theories can be represented by a graph or ``theory space'' consisting
of sites and links
\cite{Arkani-Hamed:2001ed}. In some cases, at low energies the link fields
become non-linear sigma model Goldstone fields, which are eaten to yield a
spectrum of massless and massive gauge bosons. This spectrum may match
the Kaluza-Klein tower of a compactified higher-dimensional theory
and be phenomenologically indistinguishable from an extra dimension.
More interestingly, theory space generalizes the notion of
higher-dimensional locality. This has allowed for the construction of
purely four-dimensional models that reproduce apparently higher-dimensional
mechanisms in a simple context. It has also produced powerful new
tools for physics beyond the standard model with no higher-dimensional
interpretation whatsoever.

So far, gravity has been added to these theories simply by minimally
coupling four-dimensional Einstein gravity to the four-dimensional fields. 
This is a
consistent thing to do, but then gravitational interactions do not
respect locality in theory space. In order to have gravitational
interactions be local, we would like to make gravity propagate in theory
space. This is what we consider in this paper. 
We should emphasize that we are only
interested in a low-energy description with sites and Goldstone link
fields, which in a unitary gauge reproduces a finite spectrum of massless
and massive gravitons. 
This is not ``deconstruction'', in the sense that we
are not, for the moment, interested in a full UV completion of these
theories. 
Instead, we are only trying to
make sense of gravity living in discrete spaces, and will be content with
understanding the structure of the low-energy effective theory
describing such a scenario. 

If we try to put gravity into theory space following the gauge theory example,
we have no trouble at the level of the sites.
We can easily endow each site with its own metric and general coordinate
invariance symmetry. 
The difficulty arises when we try to write down the analog of
the Goldstone boson link fields which must
transform non-trivially under two general coordinate invariances.
For instance, a four-dimensional field can only depend on one set of 
coordinates, 
not two, so how can it transform under two general coordinate invariances?
Moreover, in a unitary gauge where the links are eaten, we have a
massless diagonal graviton and a finite tower of massive gravitons.
Such theories have many strange properties, and there are doubts in the
literature about whether they are consistent
\cite{Boulware:zf,Duff:ea,Aragone:bm,Nappi:1989ny,Buchbinder:1999ar}.

In this paper, we show that there actually
is a very natural way of introducing link fields. It allows us to show
that the theory of a massless graviton coupled to a finite number of
massive gravitons makes sense as a consistent effective field theory
valid up to energies parametrically above the particle masses. 
The construction of the Goldstone link fields 
for gravity is easiest to understand
in analogy with the gauge theory case which we review in detail. 
For example, right
multiplication by an element of a gauge group 
in gauge theory translates to composition with a coordinate transformation
in the gravity case. 
This allows
us to define links, with the quantum numbers of four-dimensional vectors.
These links have simple 
transformation laws under pairs of general coordinate invariances, and
allow the construction of interacting Lagrangians with multiple general
coordinate invariances.

After describing the general formalism, we study the case of a single 
graviton of mass $m_g$ in detail. This
can be understood as a two-site model in the limit where the Planck scale on one of the
sites is taken to infinity. Just as for massive gauge theories, the
Goldstone description is extremely useful in understanding the properties
of the longitudinal polarizations of the graviton. Massive gravitons are
well-known to have a number of strange properties. The mass term must have
a specific Fierz-Pauli structure
\cite{Fierz:1939ix, VanNieuwenhuizen:fi}.
The propagator around flat space suffers
from the famous 
van Dam-Veltman-Zakharov (vDVZ) discontinuity
\cite{vanDam:1970vg,zakharov}, 
though this disappears in
anti-de Sitter (AdS) backgrounds
\cite{Karch:2001jb, Porrati:2000cp,Kogan:2000uy, Kogan:2001ub, Kogan:2001qx}.
Finally, there is the observation first made Vainshtein 
\cite{Vainshtein:sx} 
and further explored recently in
\cite{Porrati:2002cp, Deffayet:2001uk, Lue:2001gc}, 
that the gravitational field outside of a massive object breaks down at 
a peculiar macroscopic distance scale much larger than the naive gravitational
radius. We will understand all of these properties in a
transparent way, and see that they are different reflections of a single
underlying cause: the scalar longitudinal mode of the graviton becomes
strongly coupled at far lower energies than naively expected. Nevertheless,
we will see that we have a sensible effective field theory valid up to a
cutoff $\Lambda$ parametrically higher than $m_g$, with 
$\Lambda \sim (m_g^4 \mpl)^{1/5}$ for the Fierz-Pauli theory. 
This cutoff can be pushed up to
$\Lambda \sim (m_g^2 \mpl)^{1/3}$ in certain non-linear extensions of the
Fierz-Pauli Lagrangian. 
We will also be able to determine, by simple power-counting,
the form of all the corrections to the
Fierz-Pauli Lagrangian which are generated radiatively.

With the assurance that effective theories with multiple interacting spin
two fields make sense, many 
potential applications can be envisioned. The most obvious application is
to building gravitational extra dimensions, which we will discuss in an
upcoming paper. We will conclude this paper by discussing
a number of other possible applications, as well as  open problems and other directions for future
research along these lines. 

\section{Review of gauge theory} 
In this section we review some aspects of the effective 
theory of massive spin one
fields. We emphasize advantages of the effective field theory 
formalism\cite{Coleman:sm, Callan:sn}
and show
why introducing a ``fake'' gauge invariance can be a very useful thing to do.
 
Consider for definiteness an $SU(n) \times SU(n)$ gauge theory, which is
broken to the diagonal subgroup. At energies beneath the Higgsing scale, the
theory consists only of a massive and massless gauge multiplet, or
equivalently of two gauge multiplets together with the Goldstone bosons which
are eaten to make the massive gauge boson heavy. In theory space
language, we have a two-site model with gauge symmetry $SU(n)_1 \times
SU(n)_2$, with a single bi-fundamental
non-linear sigma model link field $U$:
\begin{equation}
%
%
\parbox{25mm}{
\begin{fmfgraph*}(70,40)
\fmfleft{p}
\fmfright{q}
\fmf{fermion,label=$U$,label.side=left}{p,q}
\fmfv{decoration.shape=circle,decoration.filled=empty}{p}
\fmfv{decoration.shape=circle,decoration.filled=empty}{q}
\end{fmfgraph*} }  
\end{equation}
$U$ transforms
linearly under the gauge symmetries as $U \to g_2^{-1} U g_1$ and
induces non-linear transformations on the Goldstone boson fields
when we expand $U=e^{i \pi}$.
The Lagrangian can be written as 
\begin{equation}
{\cal L} = - \frac{1}{g_1^2} \mbox{tr} F_1^2 - \frac{1}{g_2^2} \mbox{tr} F_2^2
+ f^2 \mbox{tr} |D_\mu U|^2 + \cdots
\end{equation}
where $D_\mu U = \partial_\mu U + i A_{1 \mu} U - i U A_{2 \mu}$ is the
covariant derivative. In the unitary gauge, we can set $U=1$ and the
Lagrangian becomes 
\begin{equation}
{\cal L} = - \frac{1}{g_1^2} \mbox{tr} F_1^2 - \frac{1}{g_2^2} \mbox{tr} F_2^2
+ f^2 \mbox{tr} (A_1 - A_2)^2 + \cdots
\end{equation}
In the limit as we take, say, $g_1 \to 0$, the surviving massless gauge boson
becomes all $A_1$ and completely decouples, 
and we are left with a single massive gauge boson $A \equiv
A_2$. In this limit, the Lagrangian is simply
\begin{equation}
{\cal L} = - \frac{1}{g^2} \mbox{tr} F^2 + f^2 \mbox{tr} |D_\mu U|^2 + \cdots \label{laglimU}
\end{equation}
which in unitary gauge becomes 
\begin{equation}
{\cal L} = - \frac{1}{g^2} \mbox{tr} F^2 + f^2 \mbox{tr} A^2 + \cdots \label{laglimnoU}
\end{equation}
So we can identify the mass of the gauge boson as $m_A = g f$. 

Now, the physics described by the the Lagrangian with the Goldstone bosons
included is identical to the unitary gauge Lagrangian without the Goldstone
fields. The unitary gauge Lagrangian \eqref{laglimnoU} does not have any gauge invariance,
while the Goldstone boson Lagrangian \eqref{laglimU} does.
It is clear that this
symmetry is a complete fake; we can always go to the unitary gauge where it
is not there! This is {\it always} true for local symmetries -- they are not
symmetries but redundancies of description. If a theory is
not gauge invariant, we can introduce Goldstone fields with appropriate
transformation properties to make it gauge invariant. 

However, there are important advantages to introducing the Goldstone bosons:
at energies far above $m_A$, 
the Goldstones ($\pi$) become the longitudinal component of the massive gauge
boson ($A^L$):
\begin{equation}
%
%
\parbox{25mm}{
\begin{fmfgraph*}(70,40)
\fmfleft{p1,p2}
\fmfright{q1,q2}
\fmf{photon}{p1,p}
\fmf{photon}{p2,p}
\fmf{photon}{p,q1}
\fmf{photon}{p,q2}
\fmfblob{0.3w}{p}
\fmfv{l=$A^L$,l.a=120,l.d=.05w}{p1}
\fmfv{l=$A^L$,l.a=-120,l.d=.05w}{p2}
\fmfv{l=$A^L$,l.a=60,l.d=.05w}{q1}
\fmfv{l=$A^L$,l.a=-60,l.d=.05w}{q2}
\end{fmfgraph*} }  
\quad \quad 
\overset{E \gg m_A}{\longrightarrow}
\quad \quad
%
%
\parbox{25mm}{
\begin{fmfgraph*}(70,40)
\fmfleft{p1,p2}
\fmfright{q1,q2}
\fmf{dashes}{p1,p}
\fmf{dashes}{p2,p}
\fmf{dashes}{p,q1}
\fmf{dashes}{p,q2}
\fmfblob{0.3w}{p}
\fmfv{l=$\pi$,l.a=120,l.d=.05w}{p1}
\fmfv{l=$\pi$,l.a=-120,l.d=.05w}{p2}
\fmfv{l=$\pi$,l.a=60,l.d=.05w}{q1}
\fmfv{l=$\pi$,l.a=-60,l.d=.05w}{q2}
\end{fmfgraph*} }  
\end{equation}
From the Goldstone description it is obvious that the
interactions of these longitudinal modes become strongly coupled at a scale 
$\sim 4 \pi f \sim 4 \pi m_A/g$, since this is dimensionful scale that appears
in the non-renormalizable, non-linear sigma model. 
Since the physics is exactly that of the
unitary gauge theory, this could also have been inferred directly 
in unitary gauge, though the analysis would be more cumbersome and less
illuminating. For instance, we can evaluate the Feynman diagrams for
tree-level longitudinal gauge boson scattering. Since the polarization
vector for the longitudinal gauge boson is $\epsilon_\mu \sim k_\mu/m_A$ at high
energy, there is a danger that these amplitudes can become large. 
The tree-level amplitude for $A^L A^L \to A^L A^L$ could grow
as rapidly as   
$g^2 (E/m_A)^4$ from this consideration, since there are four polarization
vectors.  However, there is a cancellation between the direct 4-point
gauge interaction and exchange diagrams for this process,  and 
the amplitude only grows as 
$\sim g^2 (E/m_A)^2$, which
becomes strongly coupled at $\sim 4 \pi m_A/g \sim 4 \pi f$. 
This example illustrates why the Goldstone boson formalism is so
powerful: it focuses on precisely the degrees of freedom that
limit the regime of validity of the effective theory by 
becoming strongly coupled. These degrees of freedom are obscured in the unitary gauge. 

The Goldstone description also allows us to determine the structure of the
higher-order terms in the effective Lagrangian by simple power-counting. 
We expect at quantum level that in addition to the leading two-derivative
terms in the non-linear sigma model, we generate higher derivative terms
such as 
\begin{equation}
\sim \frac{1}{16 \pi^2} \mbox{tr} |D_\mu U|^4, \frac{1}{16 \pi^2} \mbox{tr}
|D^2 U|^2, \cdots
\end{equation}
In unitary gauge these correspond to operators of the form 
\begin{equation}
\frac{1}{16 \pi^2} \mbox{tr} A^4, \frac{1}{16 \pi^2} \mbox{tr} (\partial A)^2, \cdots
\end{equation}

However, the natural size for these operators would have been
hard to understand directly in unitary gauge. These operators all explicitly
break gauge invariance. How would we know how to organize the
various terms that break gauge invariance? Why, for instance, 
since the theory is not gauge
invariant, do we still use the gauge invariant $F_{\mu \nu}^2$ kinetic
term but only break gauge invariance through the mass term? Normally,
this is justified because a general kinetic term gives rise to ghosts in the
propagator. But surely we must expect that non-gauge invariant kinetic terms
are generated. The standard power-counting analysis shows that non-gauge invariant
kinetic terms $\sim (\partial A)^2$ are generated, as well as non-gauge
invariant $A^4$ type interactions. But these are down by weak-coupling
factors $\sim g^2/(16 \pi^2)$ relative to the leading terms, as long as the
theory is treated as an effective theory with cutoff $\sim 4 \pi m_A/g$. These
sizes insure that all the dangerous effects associated with the non-gauge
invariant propagators, such as the appearance of ghosts, are deferred
to the cutoff, parametrically above $m_A$.  
Furthermore, it is easy to determine the natural size of all
other gauge-violating operators effects in a systematic way. 

Thus, while the Goldstone description is {\it physically identical} to the unitary
gauge description of a massive gauge boson, it is vastly more
powerful in elucidating the physics. The Goldstone description makes
it clear that the theory of a massive gauge boson is sensible up to a
cutoff $\sim 4 \pi m_A/g$. Above this scale, an ultraviolet completion is
needed. This is actually where the Goldstone description gains its full power,
since finding a UV completion only necessitates finding a good UV theory that
leads to the Goldstones at low energies. This can be done straightforwardly,
for instance via linear sigma models (as in the standard Higgs mechanism), 
or QCD-like completions (as in technicolor). But none of these UV completions
could have easily been guessed from the unitary gauge picture. 

Summarizing, the Goldstone boson description offers three advantages to
thinking about theories with massive gauge bosons: (1) 
It transparently encodes the interactions of the longitudinal
components of the gauge bosons at high energy, and determines the cutoff of
the effective theory. (2) Simple 
power-counting determines the natural size of all non-gauge invariant 
operators. (3) 
It helps point the way, from the bottom-up, to possible UV
completions of the physics.

\section{Building blocks for gravity in theory space}

We would like to do the same kind of analysis for gravity. 
We will begin with the building blocks for a general theory with many
``sites'' endowed with 
different 4D general covariances. We will show how to define ``link''
fields with suitable non-linear transformation properties, in analogy with
the gauge theory case. We will also discuss the gravitational analog of
``plaquette'' operators, needed to realize higher-dimensional theory
spaces. 

\subsection{Sites and Links}
We start with a collection of sites $j$ each of which has its own
general coordinate invariance symmetry \GC$_j$.  We will label a
given set of coordinates on the site $j$ with $x_j$. The \GC $_j$ symmetry
is generated by $x_j^\mu \to f_j^\mu (x_j)$, where we assume 
the functions $f_j$ are smooth and invertible. In this paper we will only
be concerned with local physics, and therefore ignore any issues relating
to the global topology of the sites. 

We would now like to introduce link fields that allow us to compare
objects on different sites, which transform under different \GC \,
symmetries. In order to do this in a transparent way completely parallel
to the gauge theory example, let us discuss transformation properties
under \GC \, symmetries with a simple notation. A field $\phi(x)$ is a scalar
if it transforms under \GC \,  as
\begin{equation}
\phi(x) \to \phi^\prime(x) = \phi(f(x))
\end{equation}
We can write this in a more suggestive way in terms of functional
composition:
\begin{equation}
\phi \to \phi \circ f
\end{equation}
Similarly a vector field $a_\mu(x)$ transforms under \GC \,  as
\begin{equation}
a_{\mu} (x) \to \frac{\partial f^\alpha}{\partial x^\mu}(x) a_
\alpha(f(x))
\end{equation}
We can write this more compactly by treating $a$ as a 
form: $\ba = a_{\mu}(x) {\mathbf dx}^\mu$. Then we just write
\begin{equation}
\ba \to \ba \circ f
\end{equation}
where it is understood that ${\mathbf dx}^\mu \to {\mathbf df}^\mu =
\partial_\alpha f^\mu {\mathbf dx}^\alpha$. 
This clearly generalizes to all tensor fields. For example, the transformation
properties of the metric $g_{\mu \nu}(x)$ are encoded in 
$\bg = g_{\mu \nu}(x) dx^\mu \otimes dx^\nu$ 
as $\bg \to \bg \circ f$.
Written in this way, coordinate transformations look very similar to gauge
transformations on fields in a gauge theory, e.g. $\phi^\dagger \to
\phi^\dagger g$, except we have composition in the place of group
multiplication.

Now, suppose we have two different sites $i,j$, with two different
general coordinate invariances \GC$_{i,j}$. We would like to be able to compare
fields on the two sites, which are charged under the different 
groups.
In the gauge theory case, this is accomplished by introducing
a link field $U_{ji}$ which transforms under gauge transformations as
$U_{ji} \to g_j^{-1} U_{ji} g_i$. The object $(\phi_j^\dagger U_{ji})$
transforms only under $g_i$ as $(\phi_j^\dagger U_{ji}) \to (\phi_j^\dagger
U_{ji}) g_i$. Similarly, in our case, we introduce a link field $Y_{ji}$
which transforms under \GC$_i  \times$ \GC$_j$ as
\begin{equation}
Y_{ji} \to f_j^{-1} \circ Y_{ji} \circ f_i
\end{equation}

\begin{figure}[t]
 \centerline{\epsfxsize=2 in \epsfbox{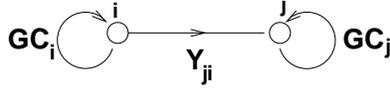}  }
\noindent
\caption{Sites $i$ and $j$ with corresponding \GC$_i \times$ \GC$_j$
symmetries, connected by a link $Y_{ji}$. Under \GC$_i \times$ 
\GC$_j$, $Y_{ji} \to f_j^{-1} \circ
 Y_{ji} \circ f_i$.}
\label{fig1}
\end{figure}

Explicitly, the link field $Y^\mu_{ji}(x_i)$ is a mapping
from the site $i$
to the site $j$ (see figure \ref{fig1}). It
associates a point on the site $i$ with coordinate $x^\mu_i$ with a point on
the site $j$ with coordinate $Y^\mu(x_i)$; and under \GC$_i \times$
\GC$_j$, we have
\begin{equation}
Y^\mu_{ji}(x_i) \to (f_j^{-1})^\mu(Y_{ji}(f_i(x_i)))
\end{equation}
$Y_{ji}$ is a pullback map from site $j$ to site $i$.

It is now clear how we can compare fields on different sites using the link
fields. 
Suppose we have a field $\psi_i$ on site $i$ and a field $\psi_j$ on site $j$
which transform under \GC$_i$ and \GC$_j$ respectively as: 
\begin{equation}
\psi_i \to \psi_i \circ f_i, \quad
\psi_j \to \psi_j \circ f_j, \quad
\end{equation}
We can construct an object $\Psi$ out of $\psi_j$ which transforms under \GC$_i$
by forming
\begin{equation}
\Psi = \psi_j \circ Y_{ji} 
\Rightarrow
\Psi \to \Psi \circ f_i
\end{equation}
Let us see how this works explicitly for various tensor fields. Starting
with a scalar $\phi_j(x_j)$ under \GC$_j$, we can form the field
\begin{equation}
\Phi(x_i) = \phi_j(Y_{ji}(x_i))
\end{equation}
which transforms as a scalar under \GC$_i$. Similarly, out of a vector
$a_{j \mu}(x_j)$ or a metric $g_{j \mu \nu}(x_j)$, we can form the objects
\begin{equation}
A_{\mu}(x_i) = \frac{\partial Y^\alpha}{\partial x_i^\mu}(x_i) a_{j
\alpha}(Y_{ji}(x_i)), 
\quad G_{\mu \nu}(x_i) = 
\frac{\partial Y^\alpha}{\partial x_i^\mu} (x_i) 
\frac{\partial Y^\beta}{\partial x_i^\nu} (x_i) 
g_{j \alpha \beta}(Y_{ji}(x_i)) \label{defG}
\end{equation}
which transform respectively as a vector and metric under \GC$_i$.

Note also that these expressions have the structure of induced tensors
familiar from brane dynamics. We can 
view the site $i$ as a spacetime filling ``brane'' embedded in the
world $j$. $Y_{ji}(x_i)$ is the location of a given point $x_i$
on the brane, in the coordinates of the spacetime in which the brane is
embedded. 

The important point is that
we can low write down a Lagrangian like:
\begin{equation}
{\cal L} =\sqrt{g_i}g_i^{\mu\nu}(g_{i \mu\nu}-G_{\mu\nu})
g_i^{\rho\sigma}(g_{i \rho\sigma}-G_{\rho\sigma}) \label{samplag}
\end{equation}
which is invariant under both GC$_i$ and  GC$_j$. We can often fix a
``unitary gauge'' where $Y$=id and this Lagrangian becomes:
\begin{equation}
{\cal L} =\sqrt{g_i}g_i^{\mu\nu}(g_{i \mu\nu}-g_{j \mu\nu})
g_i^{\rho\sigma}(g_{i \rho\sigma}-g_{j \rho\sigma}) \label{samplagU}
\end{equation}
which is exactly what we need to create mass terms for gravitons.
(Lagrangians of the form \eqref{samplagU} have been considered in 
\cite{Damour:2002ws,Damour:2002wu}.)
Introducing the $Y's$ has allowed us to introduce these mass
terms in a way that is formally fully generally covariant. Just as in the
analysis of the massive gauge boson, introducing the $Y's$ with this
``fake'' general covariance will nevertheless be extremely useful in
understanding the structure of the theory. 

\subsection{Explicit Goldstone boson expansion}

While the above construction is somewhat abstract, it is straightforward to
expand $Y$ and $G$ in terms of pions and see how the two general coordinate
invariances are realized explicitly. 
The unitary gauge has $Y$=id, that is
$Y^\mu_{ji}(x_i) = x^\mu_i$. The transformations that leave
$Y_{ji}$ = id are the diagonal subgroup of \GC$_i \times$ \GC$_j$, where
$f_i = f_j$. Even in situations with many $Y$ fields where we cannot
gauge fix all the $Y$'s to the identity, it is useful to expand the $Y'$
around a common $x$ as this corresponds to the small
fluctuations around a common background space. We therefore expand $Y$ as
\begin{equation}
Y^\alpha(x) = x^\alpha + \pi^\alpha(x)
\end{equation}
where here and in what follows we have dropped the $ij$ indices on the $Y$
and the $i$ index on $x$ to avoid notational clutter. 

Then, the object $G_{\mu\nu}$, from \eqref{defG}, can be expanded as:
\be
G_{\mu \nu} &=& \frac{\partial Y^{\alpha} ( x )}{\partial x^{\mu}}
\frac{\partial Y^{\beta} ( x )}{\partial x^{\nu}} g^j_{\alpha \beta} ( Y ( x
) )
= \frac{\partial ( x^{\alpha} + \pi^{\alpha} )}{\partial x^{\mu}}
\frac{\partial ( x^{\beta} + \pi^{\beta} )}{\partial x^{\nu}} g^j_{\alpha
 \beta} ( x + \pi ) \nn\\
&=& ( \delta_{\mu}^{\alpha} + \pi_{, \mu}^{\alpha} ) ( \delta_{\nu}^{\beta
} + \pi_{, \nu}^{\beta} ) ( g^j_{\alpha \beta} + \pi^{\mu}
g^j_{\alpha \beta , \mu} + \frac{1}{2} \pi^{\mu} \pi^{\nu} g^j_{\alpha
\beta , \mu , \nu} + \cdots ) \nn\\
&=& g^j_{\mu \nu} + \pi^{\lambda} g^j_{\mu \nu , \lambda} + \pi_{,
\mu}^{\alpha} g^j_{\alpha \nu} + \pi_{, \nu}^{\alpha} g^j_{\alpha \mu} +
\frac{1}{2} \pi^{\alpha} \pi^{\beta} g^j_{\mu \nu , \alpha , \beta} \nn\\
&&\quad\quad+
\pi_{, \mu}^{\alpha} \pi_{, \nu}^{\beta} g^j_{\alpha \beta} + \pi_{,
\mu}^{\alpha} \pi^{\beta} g^j_{\alpha \nu , \beta} + \pi_{, \nu
}^{\alpha} \pi^{\beta} g^j_{\mu \alpha , \beta}+ \cdots 
\label{Gexp}
\ee
Now we will look at the transformation properties of $g$, $G$, $Y$ and
$\pi$, under infinitesimal general co-ordinate transformations generated by
$f_i(x) = x +  \zeta_i(x)$
and $f_j(x) = x + \zeta_j(x)$. The metrics
on the sites transform as:
\be
\delta g_{\mu \nu}^i &=& \zeta_i^{\lambda} g^i_{\mu \nu , \lambda} +
\zeta_{i , \mu}^{\lambda} g^i_{\lambda \nu} + \zeta_{i ,
\nu}^{\lambda} g^i_{\mu \lambda} \label{dgi} \\
\delta g_{\mu \nu}^j &=& \zeta_j^{\lambda} g^j_{\mu \nu , \lambda} +
\zeta_{j , \mu}^{\lambda} g^j_{\lambda \nu} + \zeta_{j ,
\nu}^{\lambda} g^j_{\mu \lambda} \label{dgj} \\
\delta \sqrt{g_i} &=& \zeta^\lambda_{i, \lambda} \sqrt{g_i}+
\zeta_i^\lambda (\sqrt{g_i})_{,\lambda}, \quad
\delta \sqrt{g_j} = \zeta^\lambda_{j, \lambda} \sqrt{g_j}+
\zeta_j^\lambda (\sqrt{g_j})_{,\lambda} \label{dgg}
\ee
Thus a Lagrangian like
\begin{equation}
{\cal L} = \sqrt{g_i} R(g_i) + \sqrt{g_j} R(g_j)
\end{equation}
trivially has two general coordinate invariances. If we make the replacements
\eqref{dgi}-\eqref{dgg}, with independent $\zeta_i$ and $\zeta_j$
this Lagrangian is unchanged.
Note, however, that the ``hopping'' Lagrangian in 
\eqref{samplagU} is only invariant under the diagonal subgroup for which 
$\zeta_i = \zeta_j$. 

Now, the transformation laws of the pions come from transformation
of the link $Y$. First, under \GC$_i$:
\be
Y ( x )
\rightarrow Y ( x' ) &=& x + \zeta_i + \pi ( x + \zeta_i ) \equiv
x + \pi + \delta \pi \nn\\
\Rightarrow \delta \pi^{\mu} &=& \zeta_i^{\mu} + \zeta_i^{\alpha}
\pi_{, \alpha}^{\mu}
\ee
Under GC$_j$
\be
Y &\rightarrow& Y - \zeta_j ( Y ) = x + \pi - \zeta_j ( x + \pi )
\equiv x + \pi + \delta \pi \nn \\
\Rightarrow \delta \pi^{\mu} &=& - \zeta_j^{\mu} ( x + \pi ) 
= - \zeta_j^{\mu} - \pi^{\alpha} \zeta_{j, \alpha}^{\mu} 
- \frac{1}{2} \pi^{\alpha} \pi^{\beta} \zeta^{\mu}_{j, \alpha , \beta} + \cdots
\ee
So the pions transform under the two transformations as:
\be
\delta \pi^{\mu} &=& \zeta^{\mu}_i + \zeta_i^{\beta} \pi_{,
\beta}^{\mu} - \zeta_j^{\mu} - \pi^{\beta} \zeta_{j ,
\beta}^{\mu} - \frac{1}{2} \pi^{\alpha} \pi^{\beta} \zeta_{j ,
\alpha , \beta}^{\mu} - \cdots \label{dgp}
\ee
Note that in the global symmetry limit , 
where the $\zeta$'s are constant, 
we have
\begin{equation}
\pi^\mu \to \pi^\mu + \zeta^\nu_i  \pi^\mu_{,\nu} + \zeta_i - \zeta_j
= \pi(x+\zeta_i) + \zeta_i -\zeta_j
\end{equation}
This is just a translation in $x_i$ by $\zeta_i$, together with a
shift symmetry. Note that in this global limit the symmetry is Abelian. 
The shift symmetry is the 
analog of the shift symmetry acting on scalar Goldstone
bosons that keeps them exactly massless.

The transformations of the the pions \eqref{dgp} are 
non-linear and messy. But $G_{\mu\nu}$
has simple transformation properties which come from the simple transformations
of $Y$. By plugging \eqref{dgj} and
\eqref{dgp}
into \eqref{Gexp} we find that:
\be
\delta G_{\mu \nu} = \zeta_i^{\lambda} G_{\mu \nu , \lambda} +
\zeta_{i , \mu}^{\lambda} G_{\lambda \nu} + \zeta_{i ,
\nu}^{\lambda} G_{\mu \lambda} \label{dgG}
\ee
$G_{\mu\nu}$ transforms like a tensor under GC$_i$ and is 
invariant under GC$_j$.
This is exactly what we wanted. We can now see
that the Lagrangian \eqref{samplag} possesses two separate general coordinate 
symmetries: if we make the transformations \eqref{dgi} and \eqref{dgG}
with independent $\zeta_i$ and $\zeta_j$ the Lagrangian is unchanged. 

We can also see that is easy to make $G_{\mu\nu}$ invariant under 
both GC$_i$ and GC$_j$. 
We just redefine the non-linear transformation laws of $\pi$ to not
depend on $\zeta_i$ by setting $\zeta_i=0$ in \eqref{dgp}. Then, 
without changing the expansion of $G$ in terms of $\pi$ and $g_{j}$, 
\eqref{Gexp},  $G$ will be invariant for any $\zeta_i$ and $\zeta_j$.
This is a handy way to add general coordinate invariance to {\it any} 
Lagrangian, even one without a symmetry under the diagonal subgroup.
Again, this symmetry is a complete fake, but it can still be useful.

\subsection{Plaquettes}
In the gauge theory case, higher-dimensional theory spaces can be
constructed with the appropriate mesh of sites and links. In the case
of more than one extra dimension, not
all of the link fields can be gauged away. The classical 
spectrum includes 
the usual KK tower of spin one particles, but also a large number of
massless scalars corresponding to the uneaten link fields. It is possible
to give mass to these extra massless fields 
through the addition of ``plaquette'' interactions. For example,
going in a closed circle of links from $i$ to $j$ to $k$ back to $i$, we
can form the plaquette $(U_{ij} U_{jk} U_{ki})$, which is conjugated under
the action of $g_i$. Taking the trace the yields an invariant potential
that can be added to the Lagrangian.

We can construct plaquettes in the gravitational case as well. Suppose we have
three sites $i,j,k$ with links $Y_{ij},Y_{jk},Y_{ki}$. We can form the
functional product
\begin{equation}
\Psi = Y_{ij} \circ Y_{jk} \circ Y_{ki}
\end{equation}
which transforms as
\begin{equation}
\Psi \to f_i^{-1} \circ \Psi \circ f_i
\end{equation}

\begin{figure}[t]
 \centerline{\epsfxsize=3 in \epsfbox{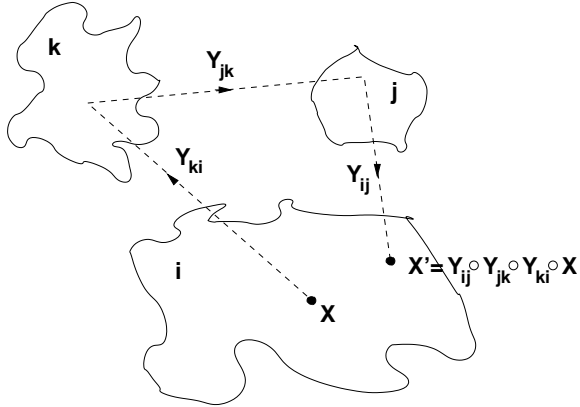}}
\noindent
\caption{Making a plaquette from $Y_{ji} \circ Y_{ik} \circ Y_{kj}$.}
\label{fig2}
\end{figure}

How can we build an invariant out of this quantity that can be added to the
action? What is the analogue of the trace in the gauge theory case?
Note that $\Psi$ maps a point $X$ on site $i$ to a different
point $X'=\Psi(X)$ (see figure \ref{fig2}). Then
the geodesic distance between $X$ and $X'$ is coordinate independent
and transforms like a scalar under \GC$_i$. 
The simplest invariant we can build which is analytic in the fields is then:
\begin{equation}
\int d^4 x \sqrt{g_i} \, l^2(x,\Psi(x);g_i) 
\end{equation}
where $l[x,y;g] = \int_x^y ds$ 
is the geodesic distance between the points $x,y$ with the
metric $g$. (We are imagining that all the $\pi$'s are perturbatively close
to zero, so that there is no ambiguity in which is the shortest geodesic
between $x$ and $\Psi(x)$.)  
This is the analog of the simplest single trace plaquette
operator in the gauge theory. Expanding around flat space, this operator
gives a mass to the uneaten combination of Goldstones $\pi^\alpha_{ij} +
\pi^{\alpha}_{jk} + \pi^\alpha_{ki}$ and provides
additional non-linear interactions needed to preserve the general 
covariances. This is all in complete analogy with the gauge theory case. 

\section{Massive gravitons}
We will now show how the effective field theory formalism
of the previous section makes studying a massive graviton
embarrassingly easy.
These fields have been sporadically
studied for many years, and are known to have 
several peculiar properties. Almost all
of the studies have been based on deforming GR by the addition of the
(already somewhat peculiar) Fierz-Pauli mass term,
$(h_{\mu \mu})^2 - h_{\mu \nu} h_{\mu \nu}$,
where $h_{\mu\nu}=g_{\mu\nu}-\eta_{\mu\nu}$ is the metric linearized
around flat space. This specific linear combination is needed
for a unitary propagator
\cite{Fierz:1939ix, VanNieuwenhuizen:fi}.
Then there is the famous van Dam-Veltman-Zakharov discontinuity
\cite{vanDam:1970vg,zakharov}
in the graviton propagator as $m_g \to 0$, which seems to indicate that
the
an arbitrarily small mass graviton yields different predictions than
Einstein's theory. Recently, it has been observed 
\cite{Karch:2001jb, Porrati:2000cp, Porrati:2002cp,
Kogan:2000uy, Kogan:2001ub, Kogan:2001qx}.
that this discontinuity
disappears in Anti-de-Sitter space and in de Sitter space, though there is
a ghost instead in the de Sitter theory. Finally, there is the observation of
Vainshtein \cite{Vainshtein:sx}, that the
discontinuity may not be relevant for physical sources, because the
linearized approximation to gravity outside a source of mass $M$ breaks   
down
at a much larger distance than the gravitational radius $R_g = l_{Pl}^2 M$;
at
a distance $(m_{g}^{-4} R_g)^{1/5}$.

With our Goldstone boson description, we will understand these
peculiarities trivially, and see that they are all associated with a 
single
underlying cause. The scalar longitudinal component of the graviton
becomes
strongly coupled at a far lower energy scale than we may have expected by
analogy with the familiar gauge theory case. For a massive spin one field,
the
cutoff is $\sim m_A/g$. This would translate to a cutoff $\sim \sqrt{m_g
\mpl}$
in the gravity case. However, we will find that while there is a sensible
effective theory for interacting massive gravitons, the cutoff is
parametrically far lower than this. Beginning with the Fierz-Pauli
Lagrangian, it is $\sim (m_g^4 \mpl)^{1/5}$, while a slightly more clever
starting
point can push the cutoff higher to $\sim (m_g^2 \mpl)^{1/3}$.

\subsection{Two site model}
Following the gauge example, it is straightforward to isolate
a single massive graviton in an arbitrary background 
from a two site model.
We start with an action is of the form 
\begin{equation}
S = S_{grav} + S_{mass} 
\end{equation}
where 
\begin{equation}
S_{grav} = \int d^4 x \sqrt{-g} \left( \mpl^2 R[g] + \cdots \right) + 
\int d^4 x_0 \sqrt{-g_0} \left(-\Lambda_0 + M_0^2 R[g_0] + \cdots \right)
\end{equation}
represents the action for the gravitons on the sites. For simplicity
we have not put in a cosmological constant term on the first site. And 
\begin{equation}
S_{mass} = \int d^4 x \sqrt{-g} g^{\mu \nu} g^{\alpha \beta} 
\left( a H_{\mu \nu} H_{\alpha \beta} + b H_{\mu \alpha} H_{\nu \beta}
\right) + \cdots 
\end{equation}
denotes the ``hopping'' action that will give one combination of gravitons
a mass. Here,  
\begin{equation}
H_{\mu \nu} (x) \equiv  g_{\mu \nu}(x) - \partial_\mu Y^\alpha(x) \partial_\nu
Y^{\beta}(x) g_{0 \alpha \beta}(Y(x))
\end{equation}
We can go to a unitary gauge where $Y$=id and there is
one manifest general coordinate invariance
under which both $g$ and $g_0$ transform as tensors. The spectrum
contains one massless graviton and one massive graviton.
In the limit where we send $M_0 \to \infty$, this massless
graviton is all $g_0$ and becomes non-dynamical, and so we are left 
with a theory of a single massive graviton described by $g$, 
in a non-dynamical background geometry $g_0$. 
(W. Siegel has informed us that this way of introducing general
coordinate invariance for a massive graviton was considered in
\cite{Siegel:1993sk}. For other early work see \cite{Green:pa}.)
The Goldstone formulation
will be invaluable in elucidating the interactions of the longitudinal 
components of the massive gravitons and determining the structure of
the effective field theory.

\subsection{Linearized analysis}
Let us begin by analyzing our action to quadratic order in the fields. To
wit, $H_{\mu \nu}$ is expanded as in \eqref{Gexp} with $g_j = g_0$:
\begin{equation}
H_{\mu \nu} = h_{\mu \nu} + g_{0 \mu\alpha}\nabla^0_\nu \pi^\alpha
+ g_{0\nu\alpha}\nabla^0_\mu \pi^\alpha + \cdots 
\end{equation}
where 
\begin{equation}
h_{\mu \nu} \equiv g_{\mu \nu} - g_{0 \mu \nu}
\end{equation}
and $\nabla^0_\mu$ is the covariant derivative with the background metric
$g_{0}$. 

Our Goldstones are a vector field, which has
3 polarizations.
These are eaten by the massless graviton, which has 2
polarizations, to produce a massive graviton with 
a total of 5 physical polarizations. We can
decompose the $\pi^\alpha$ into the transverse spin one and scalar mode by
expressing 
\begin{equation}
\pi^\alpha(x) = g_{0}^{\alpha \beta}(A_\beta + \partial_\beta \phi)
\end{equation}
This allows us to introduce a fake $U(1)$ gauge symmetry under
which 
\begin{equation}
A_\beta \to A_{\beta} + \partial_\beta \Lambda, \, \phi \to \phi - \Lambda
\label{fakeU1}
\end{equation}
The graviton mass term will turn into the kinetic term for these
Goldstones. Of course, as usual 
the new fake general covariance and $U(1)$ symmetries
must be gauge fixed by the addition of suitable gauge fixing terms for
$h_{\mu \nu}$ and $A_\alpha$, for instance fixing to Feynman-like gauges
for both. The precise form of this gauge fixing
will not be relevant for our discussion. Note that as defined 
$A_\alpha$ has mass
dimension $-1$ and $\phi$ has mass dimension $-2$. 

By the usual logic leading to the equivalence theorem for gauge theories, 
we expect that the physics of $A$ and $\phi$ is that of the vector 
longitudinal ($g^{\mathrm{vL}}$) and scalar longitudinal ($g^\mathrm{sL}$) 
polarizations of the
massive graviton field, at energies much higher than the mass of the
graviton:
\begin{equation}
%
%
\parbox{25mm}{
\begin{fmfgraph*}(70,40)
\fmfleft{p1,p2}
\fmfright{q1,q2}
\fmf{gluon}{p1,p}
\fmf{gluon}{p2,p}
\fmf{gluon}{p,q1}
\fmf{gluon}{p,q2}
\fmfblob{0.3w}{p}
\fmfv{l=$g^{sL}$,l.a=120,l.d=.05w}{p1}
\fmfv{l=$g^{sL}$,l.a=-120,l.d=.05w}{p2}
\fmfv{l=$g^{sL}$,l.a=60,l.d=.05w}{q1}
\fmfv{l=$g^{sL}$,l.a=-60,l.d=.05w}{q2}
\end{fmfgraph*} }  
\quad \quad 
\overset{E \gg m_g}{\longrightarrow}
\quad \quad
%
%
\parbox{25mm}{
\begin{fmfgraph*}(70,40)
\fmfleft{p1,p2}
\fmfright{q1,q2}
\fmf{dashes}{p1,p}
\fmf{dashes}{p2,p}
\fmf{dashes}{p,q1}
\fmf{dashes}{p,q2}
\fmfblob{0.3w}{p}
\fmfv{l=$\phi$,l.a=120,l.d=.05w}{p1}
\fmfv{l=$\phi$,l.a=-120,l.d=.05w}{p2}
\fmfv{l=$\phi$,l.a=60,l.d=.05w}{q1}
\fmfv{l=$\phi$,l.a=-60,l.d=.05w}{q2}
\end{fmfgraph*} }  
\end{equation}

Let us now consider the case where the background $g_{0 \mu \nu} =
\eta_{\mu \nu}$ is flat. We can immediately notice a peculiarity that
will be at the heart of the difference the gravity and gauge theory
cases. In the global symmetry limit where we send $\mpl \to \infty$ (
equivalently set
$h_{\mu \nu} \to 0$), there is a shift symmetry on the $\pi^\alpha$ that
means that $\pi^\alpha$ only appears with a derivative acting on it. 
This implies that $\phi$ only ever appears with two derivatives,
so $\phi^2$ must have four derivatives.
Therefore, $\phi$ cannot have a normal kinetic term in this
limit. Indeed, for general coefficients $a$ and $b$ in $S_m$, $\phi$ will
have a four-derivative kinetic term as 
\begin{eqnarray}
S_{mass} &\supset& 
\int d^4 x \, 4a \phi_{,\mu,\nu} \phi_{,\mu,\nu} 
+ 4 b \, \square \phi \square \phi \\ 
&=& \int d^4 x \, 4(a+b) \, \square \phi \square \phi
\end{eqnarray}
where to get to the second line we have integrated by parts. 
In order to eliminate this pathological kinetic term, which implies
ghosts and violations of unitarity, we must choose
$a+b=0$. This is precisely the Fierz-Pauli mass term in the unitary gauge, 
\begin{equation}
{\cal L}_{FP} = f^4 \left( h_{\mu \nu} h_{\mu \nu} - h^2 \right) \label{FPc}
\end{equation}
corresponding to a graviton mass 
\begin{equation}
m_g^2 = \frac{f^4}{\mpl^2}.
\end{equation}
We will assume throughout that the graviton mass is parametrically much smaller than the Planck scale,
$m_g \ll M_{Pl}$; this is essential for any sensible effective theory. 

Note that with the Fierz-Pauli choice $a = -b$, $\phi$ has {\it no} kinetic term in the
decoupling limit. 
On the other hand, $A_\alpha$ has a perfectly healthy kinetic term $f^4
(A_{\mu,\nu} - A_{\nu,\mu})^2$ in this limit. 
Now, $\phi$ {\it does} acquire a normal two-derivative kinetic term, but
only via mixing with the graviton $h_{\mu \nu}$. Indeed, in the expansion
of $S_{mass}$, there is is a mixing term 
\begin{equation}
f^4 \left( h_{\mu \nu} \phi_{,\mu,\nu} - h \square \phi \right)
\end{equation}
It is useful to express this term in a more familiar form. 
After integrating by parts, this kinetic mixing term is  
$f^4 \phi R_\mathrm{lin}$ where 
$R_\mathrm{lin} = h_{{\mu \nu},{\mu,\nu}} - \square h$ is the Ricci
scalar
at linear order in $h$. Thus, at quadratic order in the fields the
kinetic Lagrangian is the same as 
\begin{equation}
\sqrt{g} M_{Pl}^2 (1 + m_g^2 \phi) R
\end{equation}
We can eliminate the kinetic mixing between $\phi$ and $h$ by  
by a Weyl rescaling of the
metric, which is well-known to generate a kinetic term
for $\phi$ with the correct sign. 
At quadratic order this amounts to is redefining
\begin{equation}
h_{\mu \nu} =
\hat{h}_{\mu \nu} - \eta_{\mu \nu} (m_g^2 \phi)
\end{equation}
and the induced kinetic term for $\phi$ is 
\begin{equation}
M_{Pl}^2  m_g^4 (\partial \phi)^2 = \frac{f^8}{M_{Pl}^2} (\partial \phi)^2
\end{equation}
It will be also be convenient to add a gauge-fixing term directly for
$\hat{h}$. Then at quadratic order, we have usual kinetic and mass terms
for $\hat{h}_{\mu \nu}$, and a kinetic term for $\phi$, with no kinetic 
mixing between them.

So $\phi$ does acquire a normal kinetic term, but this term disappears in
the limit where $f$ is held fixed and $\mpl$ is sent to infinity. This is 
in stark contrast with the gauge theory case, where the Goldstone scalar
kinetic term is $\sim f^2 (\partial \pi)^2$ and survives as $g \to 0$. 

Now, we certainly do not expect our choice $a = -b$ to be exactly radiatively stable at
quantum level, so let us get a better idea of what would happen if $a \neq
-b$. The $\phi$ kinetic term would have the structure in momentum space
\begin{equation}
\frac{f^8}{\mpl^2} p^2 \phi^2 + (a+b) p^4 \phi^2
\end{equation}
and this would lead to ghosts or tachyons at a momentum scale 
$p^2 \sim \frac{f^8}{(a+b) \mpl^2} \sim m_g^2 \frac{f^4}{a+b}$. 
So we see why we need to
have $(a+b) \ll f^4$, because otherwise our effective theory would break
down right around the mass of the particle $m_g$ and would be completely 
useless. This
is the analog of the reason why in gauge theory we do not include a 
$\frac{1}{g^2} (\partial A)^2$ kinetic term. As we saw, such a term {\it is} generated
at quantum level, but with a small enough coefficient so 
that its harmful effects are deferred
to the cutoff. We will see that exactly the same thing happens in our case;
with a suitable cutoff, a small $(a+b)$ is generated (along with a whole slew of other terms), but
with small enough sizes to allow the effective theory to make sense to
energies parametrically above $m_g$. 

Notice that in order to go to canonical normalization for $\phi$, we define 
\begin{equation}
\phi = \frac{\mpl}{f^4} \phi^c = \frac{1}{m_g^2 \mpl} \phi^c
\end{equation}
The $\mpl$ in the numerator implies that the interactions of $\phi$ will
become strongly coupled at an energy far beneath $f$, again in contrast
with the gauge theory case. On the other hand, the $A_{\alpha}$ kinetic
term is proportional to $\sim f^4$, and the canonically normalized field is 
\begin{equation}
A_{\alpha} = \frac{1}{f^2} A^c_{\alpha} = \frac{1}{m_{g} \mpl} A^c_{\alpha}
\end{equation}
 
These conclusions are changed in a general curved background, which
introduces another scale into the problem. 
In flat space, with the choice $a = -b = f^4$ there is no
kinetic term for $\phi$ without mixing through $h$: the contributions
to the $\phi$ kinetic term 
proportional to $a$ and $b$ cancel exactly after integrating by parts.
In a general curved background, we have instead at
quadratic order 
\begin{eqnarray}
f^4 \int d^4 x \sqrt{-g_0} \, \left( \nabla^{0}_\mu \nabla^{0}_\nu \phi
\nabla^{0 \mu} \nabla^{0 \nu} \phi -
(\nabla^0)^2 \phi (\nabla^0)^2 \phi \right) \\
= f^4 \int d^4x \sqrt{-g_0} \, \phi_{,\mu} \left[ \nabla^{0 \mu},
\nabla^{0 \nu} \right] \phi_{,\nu} \label{nabeq}
\end{eqnarray}
after integration by parts. Since the commutator of the covariant
derivatives is proportional to the Riemann tensor and is non-vanishing in a
curved background, there is an induced kinetic term for $\phi$ (and a
corresponding mass term for $A_{\alpha}$) proportional to the background
curvature. For a maximally symmetric space
like AdS the $[\nabla^{0 \mu},\nabla^{0 \mu}]$ term can
be replaced by $\frac{1}{L^2} g^{0 \mu \nu}$ in \eqref{nabeq}, 
where $L$ is AdS radius of curvature. Therefore this contribution to
the $\phi$ kinetic term is  
\begin{equation}
\frac{f^4}{L^2} (\partial \phi)^2 = \frac{m_g^2}{L^2} \mpl^2
(\partial \phi)^2  
\end{equation}
This should be compared with the kinetic term coming from mixing with $h$, $
\sim m_g^4 \mpl^2 (\partial \phi)^2$. We see that for $1/L \gg m_g$, the
new contribution to the kinetic term dominates. 
It is easy to check that it has the good
sign in AdS space and the bad sign leading to ghosts in dS space. 
Note that for $1/L \gg m_g$, , the kinetic term survives as $\mpl$ is
taken to infinity with $f$ fixed, and in this limit  
the canonically normalized $\phi$ field is 
\begin{equation}
\phi = \frac{L}{f^2} \phi^c = \frac{1}{m_g L \mpl} \phi^c \, \, 
\left(\mbox{for} \frac{1}{L} \gg m_g
\right) 
\end{equation}

\subsection{The vDVZ discontinuity in general backgrounds}
We are now in a position to easily understand the presence/absence of the 
vDVZ discontinuity in general spacetimes. Let us first work in flat space. 
The coupling of the graviton $h_{\mu \nu}$ to the energy momentum tensor is 
\begin{equation}
T^{\mu}_{\nu} h_{\mu}^{\nu} = T^{\mu}_{\nu} (\hat{h}_{\mu}^{\nu} +
m_g^2 \delta_{\mu}^{\nu} \phi) = T^{\mu}_{\nu} \hat{h}_{\mu}^{\nu} +
\frac{1}{\mpl} T \phi^c
\end{equation}
where in the last equality we have gone to canonical normalization for
$\phi$. We see that independent of the graviton mass $m_g$, there is a
coupling of the trace of the energy momentum tensor to $\phi$ with
gravitational strength. 
Thus, for $m_g\to 0$ matter and radiation would couple with different
relative strengths than they would if $\phi$ were absent, that is
if $m_g$ were strictly massless.
This is exactly the vDVZ
discontinuity, and we have traced its origin to the strongly coupled nature
of the scalar $\phi$. Despite the fact that $h_{\mu \nu}$ only has a small
admixture $m_g^2 \eta_{\mu \nu}\phi$ of $\phi$ 
which appears to go to zero as $m_g
\to 0$, because the $\phi$ kinetic term vanishes in the same limit, when 
we go to canonical normalization the scalar couples with gravitational 
strength independent of $m_g$. 

Then we can see immediately that this discontinuity vanishes in an AdS
background. There is still a coupling  
$m_g^2 \phi T$, however in the limit where
$1/L \gg m_g$, in going to canonical normalization 
$\phi = L/(m_g \mpl) \phi^c$ we obtain the coupling 
\begin{equation}
\frac{m_g L}{\mpl} T \phi^c
\end{equation}
which {\it does} vanishes as $m_g \to 0$. There is therefore no new
gravitational strength force and no vDVZ discontinuity in this limit.  

\subsection{Strong coupling scale and power-counting in the 
effective theory}

Let us now return to flat space but consider the fully interacting theory. We
are in particular interested in the interactions of the longitudinal
components of the graviton at energies far above $m_g$, 
which are the interactions of the $\phi$ and
$A_{\alpha}$. 
These are the strongest interactions and signal when the theory breaks
down.

In flat space, our expression for $H_{\mu \nu}$ in terms of the Goldstone
bosons becomes
\begin{equation}
H_{\mu \nu} = h_{\mu \nu} + \pi_{\mu,\nu} + \pi_{\nu,\mu} +
\pi_{\alpha,\mu} \pi_{\alpha,\nu}
\end{equation}
where $\pi_\mu = \eta_{\mu\nu}\pi^\nu$.
Since we are not interested in the usual helicity two
polarization, we can set $\hat{h}_{\mu \nu} = 0$. Furthermore, since 
$\hat{h}_{\mu \nu}$ and $h_{\mu \nu}$ only differ by an amount that is
$\sim 1/\mpl \phi^c \eta_{\mu \nu}$, for the amplitudes of interest which
will be getting large far beneath $\mpl$ we can simply set $h_{\mu \nu}
= 0$ in all interaction terms. Thus, to obtain the interactions for our
Goldstones, it suffices to replace everywhere 
\begin{equation}
H_{\mu \nu} \to \pi_{\mu,\nu} + \pi_{\nu,\mu} + \pi_{\alpha,\mu}
\pi_{\alpha,\nu} 
\end{equation}
The Fierz-Pauli mass term can then be seen to contain cubic and quartic
interactions for $\phi$ and $A$. The only interactions that can become
anomalously large involve $\phi$, and are schematically 
\begin{equation}
f^4 \left[(\partial^2 \phi)^3 + (\partial^2 \phi)^4 + \partial^2 \phi
\partial A \partial A \right]
\end{equation}
(Note that it is impossible to have a term with a single $\partial A$ and
$\partial ^2 \phi$'s, because by the $U(1)$ gauge invariance \eqref{fakeU1},
the $\partial A$ piece would have 
to involve $F_{\mu \nu}$ which is antisymmetric, and
vanishes when contracted with anything made out of $\phi_{,\mu, \nu}$ which
is symmetric in $\mu,\nu$.)   

Going to canonical normalization, these interactions become 
\begin{equation}
\frac{1}{m_g^4 \mpl} (\partial^2 \phi^c)^3 + \frac{1}{m_g^6 \mpl^2}
(\partial^2 \phi^c)^4 + \frac{1}{m_g^2 \mpl} \partial^2 \phi^c \partial
A^c \partial A^c 
\end{equation}
The cubic scalar interaction is the strongest coupling and
becomes large at an energy scale 
\begin{equation}
\Lambda_5 \sim (m_g^4 \mpl)^{1/5}
\end{equation}
Correspondingly, the amplitude ${\cal A}(\phi \phi \to \phi \phi)$ 
from $\phi$ exchange grows as $\sim E^{10}/\Lambda_5^{10}$ and gets strongly
coupled at $\Lambda_5$. This means that the scattering amplitude for the scalar longitudinal 
polarization of the massive graviton gets strongly
coupled at the scale $\Lambda_5$:

\begin{equation}
%
%
%
\parbox{25mm}{
\begin{fmfgraph*}(70,40)
\fmfleft{p1,p2}
\fmfright{q1,q2}
\fmf{gluon}{p1,p}
\fmf{gluon}{p2,p}
\fmf{gluon}{p,q1}
\fmf{gluon}{p,q2}
\end{fmfgraph*} }  +
%
%
\parbox{30mm}{
\begin{fmfgraph*}(85,40)
\fmfleft{p1,p2}
\fmfright{q1,q2}
\fmf{curly}{p1,pl}
\fmf{curly}{p2,pl}
\fmf{curly}{pl,pr}
\fmf{curly}{pr,q1}
\fmf{curly}{pr,q2}
\end{fmfgraph*} }  =
\parbox{30mm}{
\begin{fmfgraph*}(85,40)
\fmfleft{p1,p2}
\fmfright{q1,q2}
\fmf{dashes}{p1,pl}
\fmf{dashes}{p2,pl}
\fmf{dashes}{pl,pr}
\fmf{dashes}{pr,q1}
\fmf{dashes}{pr,q2}
\end{fmfgraph*} }
+\quad \text{smaller amplitudes} \label{gravscat}
\end{equation}
\vskip 0.1in
This could have actually been guessed directly in unitary gauge. The polarization vector for the scalar 
longitudinal polarization of the graviton at high energies is
$\epsilon_{\mu \nu} \sim k_\mu  k_\nu / m_g^2 \sim (E^2/m_g^2)$. Naively,
each one of the graviton diagrams in \eqref{gravscat} grow as 
\begin{equation}
\left(\frac{E^2}{m_g^2}\right)^4 \times \frac{E^2}{\mpl^2} \sim
\frac{E^{10}}{\Lambda_5^{10}}
\end{equation}
In the gauge theory case, there is a cancellation of the leading
term between the two diagrams; one may have expected a similar
cancellation here. However, without the need to perform this very hairy
perturbative massive gravity computation, our Goldstone description tells
us that no such
cancellation occurs, and the amplitude gets strongly coupled at
$\Lambda_5$. Perhaps this is not surprising, since our starting point, the
Fierz-Pauli Lagrangian, is somewhat arbitrary. That is,
there is nothing special about terms quadratic in $h$ given that $h$ 
is dimensionless. Higher order interactions may help cancel
the strongest divergences in \eqref{gravscat}. We will come back to this
in section \ref{secraise}.

Let us proceed to determine the structure of the operators 
generated at quantum level in this effective theory,  taken to have a cutoff
$\Lambda_5$. We must include all operators consistent with the symmetries,
suppressed by the cutoff $\Lambda_5$. The shift symmetry guarantees that
the leading operators are of the form 
\begin{equation}
\frac{\partial^q (\partial^2 \phi^c)^p}{\Lambda_5^{3p + q - 4}}
\end{equation}
In order to find what operators these correspond to in unitary gauge, we
can go back to the original normalization for $\phi^c = m_g^2 \mpl \phi$ 
and recall that
$\phi_{,\mu, \nu}$ always comes from an $h_{\mu \nu}$. Thus,
in unitary gauge, we have operators of the form 
\begin{equation}
c_{p,q} \partial^q h^p
\end{equation}
where the coefficients $c_{p,q}$ have a natural size
\begin{equation}
c_{p,q} \sim 
\Lambda_5^{-3p-q+4} \mpl^p  m_g^{2p} =
\left(m_g^{16 - 4q - 2p} \mpl^{2p - q + 4}\right)^{1/5} 
\end{equation}
Note that, for example, the term with $p=2,q=0$ is a general mass term for $h$, not
necessarily of the Fierz-Pauli form. However, its coefficient, 
$c_{2,0}=(m_g^{12/5} \mpl^{8/5})$, is parametrically 
much smaller than the Fierz-Pauli coefficient
$f^4=m_g^2 \mpl^2$ \eqref{FPc},
and the unitarity violation/tachyons are
 postponed to energies above the cutoff $\Lambda_5$. 

We can summarize by saying that there is a natural
effective theory with an action
\begin{equation}
\int d^4 x \sqrt{-g} ( \mpl^2 R + \cdots ) + 
m_g^2 \mpl^2 (h_{\mu \nu}^2 - h^2) +
\sum_{p,q} c_{p,q} \partial^q h^p
\end{equation}
with a cutoff $\Lambda_5 = (m_g^4 \mpl)^{1/5}$. The ``pollution'' from all the higher order
terms do not give rise to any pathologies until above $\Lambda_5$. Needless
to say, it would have been hard to guess the structure of this effective
theory directly from a unitary gauge analysis. 

\subsection{Adding interactions to raise the cutoff \label{secraise}}
We can easily find another natural effective theory where the cutoff is
parametrically higher than $\Lambda_5$.  By adding higher-order
terms of the form $f^4(h^3 + h^4 + \cdots)$, we can remove {\it all} the
$\phi$ self-couplings from the action. There is no unique procedure,
but one
way is as follows. Since $H_{\mu \nu} \sim \phi_{,\mu,\nu} +
O((\partial^2 \phi)^2)$, at any order we can cancel all terms of the form
$(\partial^2 \phi)^n$ by appropriately choosing the coefficient of $H^n$
terms. These in turn only generate higher order $\phi$ self-interactions,
which are canceled at the next step. Having eliminated all the $\phi$
self-interactions with terms of the form $f^4 H^n$, we are left with
interactions of the form 
\begin{equation}
f^4 (\partial A)^p (\partial^2 \phi)^q = \frac{1}{m_g^{p + 2q -2} \mpl^{p
+ q -2}} (\partial A^c)^p (\partial^2 \phi^c)^q
\end{equation}
for $p>1$. These become strongly coupled at a scale 
\begin{equation}
\left(m_g^{p + 2q -2} \mpl^{p + q -2}\right)^{\frac{1}{3q + 2p - 4}} 
\end{equation}
It is easy to see that the lowest this scale can ever be is $\Lambda_3 =
(m_g^2 \mpl)^{1/3}$, which is achieved for $p=2,q=1$, and also
asymptotically as $q \to \infty$. Therefore the cutoff of this effective theory
is $\Lambda_3$. Note that with this choice, the leading contribution to
the $\phi \phi \to \phi \phi$ amplitude is absent, which 
means that there is a partial cancellation between the two unitary gauge 
diagrams on the left side of \eqref{gravscat}. 
The largest amplitude is that of 
$A A$ scattering through $\phi$ exchange, 
where the amplitude again grows as naively expected in unitary gauge $\sim
E^6/(m_g^4 \mpl^2)$, becoming strongly coupled at $\sim \Lambda_3$. 
It is easy to find that the natural size for 
operators of the form $c_{p,q} \partial^q h^p$ in
unitary gauge is now
\begin{equation}
c_{p,q} \sim \Lambda_3^{4-q} 
=\left(m_g^2\mpl\right)^\frac{4-q}{5} 
\end{equation}
Again, the choice of coefficients needed to eliminate all the $\phi$
self-interactions is technically natural; since $\Lambda_3 \ll f =
\sqrt{m_g \mpl}$, the pollution from the operators not of the special
form is small and pathologies are postponed to the cutoff. 

This makes it tempting to try to push the cutoff higher
by adding other interactions. We will address this question
in detail elsewhere \cite{us}. 

\subsection{Breakdown of the effective theory around heavy sources}
We have seen that our effective field theory breaks down at high energy
scales $\Lambda \sim \Lambda_5$ or 
$\Lambda_3$, and there are infinitely many higher dimension operators
suppressed by $\Lambda$  
that encode our ignorance of the short-distance UV completion of these theories. 
In high-energy 
scattering experiments, these effects only become important near the cutoff scale. 
But quite generally, effective theories can also break down at large
distance scales in the presence
of large background fields that make the higher dimension operators important. 
For instance, the Euler-Heisenberg Lagrangian which describes 
electrodynamics at energies beneath the electron mass contains higher-dimension operators of the 
form $\sim F^4/m^4, F^6/m^8 \cdots$. 
The effective theory certainly breaks down at short distances
of order $m^{-1}$, but it can also break down in backgrounds with spatially homogeneous but large 
electric fields, where $F/m^2$ becomes $\sim 1$. Schwinger pair production becomes important and the 
structure of the short-distance physics becomes relevant even at large distances. 
Similarly, in our case there are higher-dimension operators which become important when 
$\partial^2 \phi^c/\Lambda^3$ becomes $\sim 1$, and this quantity can indeed become large
in the presence of heavy sources with masses $M$ much larger than $M_{Pl}$. 
 
Consider for instance the potential field for $\phi^c$. Since $\phi^c$ couples with gravitational
strength, it will affect the motion of test particles around a heavy source. 
The potential set up for $\phi^c$ by a source of mass $M$ can be diagrammatically represented as

\begin{equation}
%
%
\parbox{40mm}{
\begin{fmfgraph*}(75,40)
\fmfleft{pl}
\fmfright{pr}
\fmf{dashes}{pl,pr}
\fmfv{decoration.shape=circle,decoration.filled=hatched,decoration.size=.07width}{pl}
\end{fmfgraph*} } \quad\quad 
%
%
\parbox{40mm}{
\begin{fmfgraph*}(85,40)
\fmfleft{p1,p2}
\fmfright{pr}
\fmf{dashes}{p1,pl}
\fmf{dashes}{p2,pl}
\fmf{dashes}{pl,pr}
\fmfv{decoration.shape=circle,decoration.filled=hatched,decoration.size=.07width}{p1}
\fmfv{decoration.shape=circle,decoration.filled=hatched,decoration.size=.07width}{p2}
\end{fmfgraph*} } \quad\quad 
%
%
%
%
\parbox{40mm}{
\begin{fmfgraph*}(85,40)
\fmfleft{p1,p2,p3}
\fmfright{pr}
\fmf{dashes}{p1,pl}
\fmf{dashes}{p2,pl}
\fmf{dashes}{p3,pl}
\fmf{dashes}{pl,pr}
\fmfv{decoration.shape=circle,decoration.filled=hatched,decoration.size=.07width}{p1}
\fmfv{decoration.shape=circle,decoration.filled=hatched,decoration.size=.07width}{p2}
\fmfv{decoration.shape=circle,decoration.filled=hatched,decoration.size=.07width}{p3}
\end{fmfgraph*} } 
\end{equation}
\vskip .1in
Here the blobs denote the heavy external source.  The first 
diagram represents the potential set up for $\phi^c$ at linearized level, which will be a good approximation 
at sufficiently large distances from the source. The coupling
to the source is proportional to $M/\mpl$ so           
\begin{equation}
\phi^c{}^{(1)} \sim \frac{M}{\mpl} \frac{1}{r}  \label{vr1}
\end{equation}
The remaining diagrams can be easily estimated. 
Each vertex gives us a factor of $(M/M_{Pl})$, while from $n$ point vertices of the form 
$\partial^q (\partial^2 \phi_c)^n/\Lambda^{3n + q - 4}$ there is a factor 
of $1/\Lambda^{3n + q - 4}$. The remaining $1/r$ factors are fixed by dimensional 
analysis, and we find the contribution to $\phi_c$ is 
\begin{equation}
\phi_c^{(n,q)} \sim (\frac{M}{M_{Pl}})^{n-1} \frac{1}{\Lambda^{3n + q - 4}}
\frac{1}{r^{3n + q - 3}}
\end{equation}
The distance $r_n$ at which 
the $n$'th order contribution to $\phi_c$ becomes comparable to the lowest 
order contribution is then 
\begin{equation}
r_{n,q} \sim \Lambda^{-1} (\frac{M}{M_{Pl}})^{\frac{n-2}{3n + q - 4}}
\end{equation}
Note interestingly that this distance {\it increases} with $n$, and asymptotes to 
\begin{equation}
r_*\sim \Lambda^{-1} (\frac{M}{M_{Pl}})^{1/3}
\end{equation}
This distance is precisely where $\partial^2 \phi^{c(1)}/\Lambda^3$
becomes $\sim 1$. In the action, terms with the minimal number of
derivatives $q=0$, but with any $n$, all become important at the distance
$r_*$. 

This tells us two things. First, the effective theory around heavy sources breaks down at
parametrically much larger distances than the short distance cutoff $\Lambda^{-1}$, by a factor 
of $(M/M_{Pl})^{1/3}$, since infinitely many higher dimension operators 
become important at this scale.  Second, there is no range of
distances for which the linear approximation breaks down but non-linear effects can be reliably 
computed in
the effective theory, {\it i.e.} 
for which only a finite number of the higher dimension operators become 
important. This is because it is the 
{\it highest} dimension operators  
that contribute to the onset of 
non-linear effects at large distances. 
We therefore directly transition from the 
linear regime to one where the effective theory breaks down. This is in 
contrast with non-linear effects in Einstein Gravity. Here, at
a distance of order the gravitational radius, the linearized approximation
breaks down, and all the operators with two derivatives and any number of
$h$'s become important. However, general covariance dictates that all these
higher dimension operators are packaged together into the Ricci scalar $R$,
and therefore all their coefficients are known. We can therefore trust the
non-linear gravity solution to much smaller distances, and the effective
theory only breaks down when the curvatures become Planckian. In our
case, we don't have any symmetry 
principle analgous to general covariance to 
determine the coefficients of all operators with $q=0$ for any $n$, and
therefore without a UV completion we are unable to compute non-linear
effects around heavy sources consistently within the effective theory.  

It is interesting to compare our conclusions here with the observations of 
Vainshtein 
\cite{Vainshtein:sx}
that foreshadowed some of these results. Working with the 
Fierz-Pauli theory, purely at the classical level,
Vainshtein found that the linear
approximation for gravity breaks down at a macroscopic distance 
$r_V \sim (G_N M m_g^{-4})^{1/5}$ 
from the source. He further argued that the full non-linear solution would have a continuous 
behavior as $m_g$ is taken to zero. We can understand the origin of the 
Vainshtein radius 
trivially. 
Recall that just with the Fierz-Pauli term,
the strongest interaction was the triple-scalar 
interaction suppressed by $\Lambda_5$. Therefore
the contribution $n=3,q=0$ in our analysis above is the largest one, and
the corresponding radius $r_{3,0}$, 
where this non-linear correction becomes
comparable to the lowest-order term, 
is precisely $r_V$. 
However, from the point of view of the effective theory, there is no reason to 
only keep the Fierz-Pauli terms, and in fact we must
include all other operators with their natural sizes.
Doing this we have found even larger non-linear effects. We
have seen that the entire effective 
theory breaks down at the distance $r_*$, 
so that it is impossible to make any reliable predictions 
for gravitational strength forces at distances smaller 
than $r_*$ without specifying the UV
completion of the theory. 
And we certainly do not have any reason to expect a smooth limit as $m_g$ 
is taken to zero, since in this limit the distance scale at which the
effective
theory breaks down goes to infinity. 

If we wish to consider
the possibility that our four-dimensional graviton has a small 
mass, then its Compton
wavelength should be on order of the size of the universe: 
$m_g \sim 10^{28}$ cm${}^{-1}$. 
Then, we find that $\Lambda_5^{-1} \sim$ 10$^{13}$ cm. Even if
we modify the theory to raise the cutoff to $\Lambda_3$,  we
would only have $\Lambda_3^{-1} \sim 10^{7}$ cm. The effective theory breaks down at
even larger distances around heavy sources. 
Both of these scales are far larger than $\sim 1$ mm, 
where gravitational effects have been measured. Therefore
our effective theory for a massive 4D graviton breaks down at distances
larger than the scale we have measured gravity, and we cannot say
in any controlled way that this theory is consistent with experiment.
In order to avoid conflict with experiment, the short distance cutoff must 
at least be pushed to $\sim$ mm, which is around $\sqrt{m_g M_{Pl}}$. 
We will discuss this possibility in more detail in \cite{us}.

\section{Summary, Discussion and Outlook}
We have shown how to understand massive gravitons within
the language of effective field theory. 
We are now able to write down interacting gravitational Lagrangians in a theory space
with multiple copies of general coordinate invariance. The
key is to introduce link fields which transform non-linearly under various
transformations. In unitary gauge, these links are eaten to make the
gravitons massive.
Our generally covariant formalism allows us to study the largest
interaction in the theory, involving the longitudinal components of the
massive gravitons, in a simple way. 

As an illustration, we have applied this formalism
to study a single graviton of mass $m_g$. 
We find that there is a consistent effective theory with a cutoff 
$\Lambda$ which can be taken parametrically higher than 
$m_g$. It can be taken as high
as $\Lambda \sim (\mpl m_g^4)^\frac{1}{5}$ for the simplest case based on 
the Fierz-Pauli theory, and
as high as $\Lambda \sim (\mpl m_g^2)^\frac{1}{3}$ if we add additional
terms beyond the Fierz-Pauli structure. We have understood
a number of strange features of massive gravitons in a transparent way, and 
seen that they are all consequences of the peculiar behavior of the scalar
longitudinal component of the graviton, $\phi$.
That the mass term must have Fierz-Pauli form to guarantee
unitarity follows immediately from eliminating the pathological large
four-derivative kinetic term for $\phi$.  
Having done this, around flat space $\phi$ 
only acquires a kinetic term by mixing with $h_{\mu \nu}$. 
This is the origin of the vDVZ discontinuity. However, around curved
backgrounds, such as AdS space, $\phi$ does pick up a normal kinetic term
proportional to the background curvature even without mixing, 
and therefore the vDVZ discontinuity is absent. We
have shown how to include all terms beyond the Fierz-Pauli 
Lagrangian with their natural sizes in the effective theory, 
and in particular observed that the Fierz-Pauli form of the mass term is
radiatively stable. We also saw that around sources of mass $M$ much larger
than $M_{Pl}$, the effective theory breaks down at much larger distances
than the short-distance cutoff scale, parametrically at a radius $\sim
(M/M_{Pl})^{1/3} \Lambda^{-1}$. 

Of course, the purpose of the effective field theory formalism is
not just to understand a single massive graviton.
Now that we understand the dynamics of theories
with multiple interacting gravitons,  we can construct large classes of
models with gravity in theory space.
As a trivial example, we can consider a theory
space version of the first Randall-Sundrum model
\cite{Randall:1999ee}. We have two sites, one
with TeV scale gravity and the standard
model and the other with Planck scale gravity. A simple 
link field will give a TeV mass to one combination of gravitons. 
At low energies there is a massless graviton 
with ordinary Planck scale
couplings, together with a massive graviton with 1/TeV couplings to
the Standard Model fields. Note that in our set-up there is no need to
introduce and stabilize a radion. It would be interesting to extend
the theory to more sites in a way that would dynamically generate the large
hierarchy of scales in a natural way. Nevertheless, the two-site model has
a low quantum gravity cutoff of $\sim$ TeV which cuts off the Higgs mass quadratic divergence.
More generally, we can construct models which involve gravitationally 
sequestered sectors
weakly coupled to the standard model. 
It should also be straightforward to extend our methods to understanding 
supergravity in theory space. An obvious application of these ideas would
be the communication of supersymmetry
breaking between sites. For instance we could consider the theory space
version of anomaly mediation. It will also be interesting to explore cosmological issues in few site
models. 
In addition to simple constructions with a few sites, we can also consider 
building gravitational dimensions with many sites,  which leads to some
fascinating physics that will be discussed in detail in \cite{us}.

The most interesting of all possible applications would be the
construction of a UV complete theory of gravity. Recall that
the {\it deconstruction} of non-renormalizable 
higher-dimensional gauge theories has provided them with a UV
completion. The structure of the high energy theory 
was easily guessed at by attempting to UV complete the low-energy 
non-linear sigma model fields with spontaneous symmetry
breaking. Furthermore, with the 
addition of extra ingredients, such as supersymmetry and conformal
invariance, these models lead to deconstructions of non-gravitational
sectors of string theory \cite{little}. It is a
tantalizing possibility that by pursuing the analogy with the gauge theory
which we have begun to develop in this paper, perhaps with some additional
ingredients, we may be led to UV completions of four-dimensional quantum
gravity.

\acknowledgments 
We would like to thank Andy Cohen and Thomas Gregoire for useful
discussions. We understand that M. Luty, M. Porrati, R. Rattazzi and
R. Sundrum have been exploring massive gravity in the context of the
DGP model \cite{Dvali:2000hr}. We thank them for discussions. Also,
A. Chamseddine has informed us that related ideas can be found in
\cite{Chamseddine:1977bi}.
N.A-H. is supported in part by the Department of
Energy under Contracts DE-AC03-76SF00098 and the National Science
Foundation under grant PHY-95-14797, the Alfred P. Sloan foundation,
and the David and Lucille Packard Foundation.

\end{fmffile}

\end{document}